\RequirePackage{xspace}

\def\babar{\mbox{\slshape B\kern-0.1em{\smaller A}\kern-0.1em
    B\kern-0.1em{\smaller A\kern-0.2em R}}}

\newcommand{\jprlBase}       {Phys.\ Rev.\ Lett.\xspace}
\newcommand{\jprBase}        {Phys.\ Rev.\xspace}
\newcommand{\jplBase}        {Phys.\ Lett.\xspace}

\newcommand{\npBase}         {Nucl.\ Phys.\xspace}

\newcommand{\npb}       [1]  {\npBase\ B~{\bf #1}}

\newcommand\etal{{\it et al.}}

\newcommand{\plb}       [1]  {\jplBase\ B~{\bf #1}}

\newcommand{\jprl}      [1]  {\jprlBase\ {\bf #1}}

\newcommand{\jprd}      [1]  {\jprBase\ D~{\bf #1}}

\documentclass{PoS}

\title{Measurements of charmless B decays related to $\alpha$ at BABAR}

\ShortTitle{Measurements of charmless B decays related to $\alpha$ at BABAR}

\author{\speaker{Vincenzo LOMBARDO} (for the BABAR Collaboration)\\
       INFN Milano (Italy)\\
       E-mail: \email{vincenzo.lombardo@mi.infn.it}\\
       \\
       BABAR-PROC-09/082, SLAC-PUB-13797}

\abstract{We report recent measurements of the CKM angle $\alpha$ using data collected by the BABAR detector at the PEP-II 
asymmetric-energy $e^+e^-$ collider at the SLAC National Accelerator Laboratory. In addition to improved constraints on 
$\alpha$ from the decays $B^{\pm}\rightarrow \rho^{\pm}\rho^0$, we also present preliminary results of neutral and charged 
$B$ meson decays to $K_1(1270) \pi$ and $K_1(1400) \pi$ and its impact on the estimate for the CKM angle $\alpha$ based on 
time-dependent analysis of CP-violating asymmetries in $B^0\rightarrow a_1(1260)^{\pm} \pi^{\mp}$. Moreover we report the 
first observation of the decay $B\rightarrow a_1(1260)^{\pm}a_1(1260)^{\mp}$; this mode can be used, in principle, to provide 
an indipendent measurement of $\alpha$.}

\FullConference{European Physical Society Europhysics Conference on High Energy Physics\\
   July 16-22, 2009\\
    Krakow, Poland}

\begin{document}

\section{Introduction}
\label{sec:intro}
\noindent
The measurements of the angles $\alpha$, $\beta$ and $\gamma$ of the Unitarity Triangle (UT)
at the B-factories are providing precision tests of the description of CP violation in
the Standard Model (SM). This description is provided by the Cabibbo-Kobayashi-Maskawa (CKM)
quark-mixing matrix~\cite{CKM,wolfenstein}. I am summarizing here recent experimental results
on the Unitarity Triangle angle $\alpha$ obtained from B-meson decays with the BABAR experiment 
at the SLAC National Accelerator Laboratory. The BABAR detector and PEP-II accelerator are 
described elsewhere~\cite{babar}. The decays of neutral $B$ mesons to the final states $h^{+} h^{'-}$, 
where $h^{+},h^{'-}$ = $\pi$, $\rho$, $a_1$ are sensitive to the CKM angle $\alpha$ in the 
interference between decay and mixing~\cite{a1}. The presence of gluonic loop (``penguin'') 
contributions with a different weak phase to the tree contribution shifts the measured angle 
from the UT angle $\alpha$ to an effective parameter $\alpha_{eff}$, where the shift is defined as  
$\delta \alpha = \alpha - \alpha_{eff}$. Either isospin symmetry \cite{GL,Gross} or broken SU(3) 
flavor symmetry \cite{GZ} can be employed to disentangle $\alpha$ from $\alpha_{eff}$.

\section{Measurement of B decays to $\rho^{\pm}\rho^0$}
\noindent
One of the most favorable methods to determine $\alpha$ is through an isospin analysis of the $B\rightarrow \rho \rho$ system~\cite{amsler,fitter}. 
Here we present updated BaBar results for $B^+ \rightarrow \rho^+\rho^0$ channel, with $\rho^+\rightarrow \pi^+\pi^0$
and $\rho^0\rightarrow\pi^+\pi^-$, leading to an improved determination of $\alpha$~\cite{newrho}. The analysis is based on $(465 \pm 5)\times 10^6 B\bar{B}$~events 
collected on the $Y(4S)$ resonance. Compared to the previous study~\cite{rhorhoprevious}, the analysis incorporates higher signal efficiency and 
background rejection, twice as much as data, and an improved procedures to reconstruct charged particles and to account for correlations in the 
backgrounds. The measured branching fraction ${\cal B}(B^+ \rightarrow \rho^+\rho^0) = (23.6 \pm 1.4 \pm 1.4)\times 10^{-6}$ 
is larger than Ref.~\cite{rhorhoprevious}, primarly becouse of the improved method used here to account for correlations in the backgrounds. The measured longitudinal 
polarization fraction is $f_L$ = $0.950 \pm 0.015 \pm 0.006$. An isospin analysis of $B \rightarrow \rho\rho$ has been performed by minimizing a $\chi^2$ that 
included the measured quantities expressed as the lenghts of the sides of the 
$B$ and $\bar{B}$ isospin triangles~\cite{GL}. 
The $B^+ \rightarrow \rho^+\rho^0$ branching fraction and $f_L$ results presented here have been used, with the branching fractions, polarizations and CP-violating 
parameters in $B^0 \rightarrow \rho^+\rho^-$~\cite{rho} and $B^0 \rightarrow \rho^0\rho^0$ decays reported in Ref.~\cite{rhoz}. We assume the uncertainties to be 
Gaussian-distributed and neglect potential isospin $I=1$ and elettroweak-loop amplitudes which are expected to be small. The CKM phase angle $\alpha$ and its correction 
$\Delta\alpha$ are found to be 
$\alpha = (92.4^{+6.0}_{-6.5})^{\circ}$ and $1.8^{\circ}$ < $\Delta\alpha$ < $6.7^{\circ}$, respectively at 68\% C.L., significant improvement compared to 
$\alpha = (82.6^{+32.6}_{-6.3})^{\circ}$ and $|\Delta\alpha| <15.7^{\circ}$ obtained with the same $\rho^+\rho^-$ and $\rho^0\rho^0$ measurements but the previous 
$B^+ \rightarrow \rho^+\rho^0$ results. The improvement is primarly due to the increase in ${\cal B}(B^+ \rightarrow \rho^+\rho^0)$ compared to our previous result.
${\cal B}(B^+ \rightarrow \rho^+\rho^0)$ determines the length of the common base of isospin triangles for $B$ and $\bar{B}$ decays. The increase in the base length 
flattens both triangles, making the four possible solutions nearly degenerate.

\section{Measurement of B decays to $K_1(1270)\pi$ and $K_1(1400) \pi$}
\noindent
BaBar has recently reported the measurement of branching fractions of neutral and charged $B$ meson decays to $K_1(1270)\pi$ and $K_1(1400)\pi$, 
obtained from a data sample of 454 million $\Upsilon(4S)\to B\bar{B}$ events~\cite{Simone}. The signal is modeled with a $K$-matrix 
formalism, which accounts for the effects of interference between the $K_1(1270)$ and $K_1(1400)$ mesons.
Including systematic and model uncertainties, BaBar measures 
${\cal B}(B^0\to K_1(1270)^{+}\pi^-+K_1(1400)^{+}\pi^-)= 3.1^{+0.8}_{-0.7} \times 10^{-5}$ and 
${\cal B}(B^+\to K_1(1270)^{0}\pi^++K_1(1400)^{0}\pi^+)= 2.9^{+2.9}_{-1.7} \times 10^{-5}$ ($<8.2\times 10^{-5}$ at 90\% probability).
A combined signal for the decays $ B^0\to K_1(1270)^{+}\pi^-$ and $B^0\to K_1(1400)^{+}\pi^-$ is observed with
a significance of $7.5\sigma$, and the following branching fractions are derived for neutral $B$ meson decays:
${\cal B}(B^0 \to K_1(1270)^{+}\pi^-) \in [0.6,2.5]\times 10^{-5}$,~
${\cal B}(B^0 \to K_1(1400)^{+}\pi^-) \in [0.8,2.4]\times 10^{-5}$, and 
${\cal B}(B^0 \to K_{1A}^{+}\pi^-) \in [0.4,2.3]\times 10^{-5}$,~
where the $K_{1A}$ meson is a nearly equal admixture of the $K_1(1270)$ and $K_1(1400)$ resonances~\cite{amsler} and where the 
two-sided intervals are evaluated at 68\% probability. A significance of $3.2\sigma$ is obtained for $B^+\to K_1(1270)^{0}\pi^++K_1(1400)^{0}\pi^+$,
and the following two-sided intervals at 68\% probability and 
upper limits at 90\% probability are derived:
${\cal B}(B^+ \to K_1(1270)^{0}\pi^+) \in [0.0,2.1]\times 10^{-5}$~($<4.0\times 10^{-5}$),~
${\cal B}(B^+ \to K_1(1400)^{0}\pi^+) \in [0.0,2.5)\times 10^{-5}$~($<3.9\times 10^{-5}$), and 
${\cal B}(B^+ \to K_{1A}^{0}\pi^+) \in [0.0,2.1]\times 10^{-5}$~($<3.6\times 10^{-5}$).
Moreover BaBar has combined these branching fractions with existing experimental information to derive an independent estimate for the CKM 
angle $\alpha$, based on the time-dependent analysis of $CP$-violating asymmetries in $B^0 \rightarrow a_1(1260)^{\pm} \pi^{\mp}$~\cite{ALPHA}. 
The $\Delta S = 1$ decays presented here are particularly sensitive to the presence of penguin amplitudes because their CKM couplings 
are larger than the corresponding $\Delta S = 0$ penguin amplitudes. Thus measurements of the decay rates of the $\Delta S = 1$ transitions involving the same 
SU(3) flavor multiplet as $a_1(1260)$ provide constraints on $\Delta\alpha=\alpha_{\rm eff} -\alpha$~\cite{BOUNDS}. 
Similar SU(3)-based approaches have been proposed for the extraction of $\alpha$ in the $\pi^+\pi^-$~\cite{PIPIPH}, $\rho^{\pm}\pi^{\mp}$~\cite{GZ}, and $\rho^+\rho^-$ channels~\cite{RHORHOPH,rho}.
Babar derives bounds on the model uncertainty $|\Delta \alpha|$ on the weak phase $\alpha_{\rm eff}$ extracted in $B^0\rightarrow a_1(1260)^{\pm}\pi^{\mp}$ decays using 
previously measured branching fractions of $B^0\rightarrow a_1(1260)^{\pm}\pi^{\mp}$, $B^0\rightarrow a_1(1260)^{-}K^{+}$ and $B^+\rightarrow a_1(1260)^{+}K^0$ decays~\cite{APDECAYSA1} and the $CP-$violation asymmetries~\cite{ALPHA} as input to the method of Ref.~\cite{BOUNDS} and obtains $|\alpha -\alpha_{\rm eff}|<11^{\circ}(13^{\circ})$ at 68\% (90\%) probability. The determination of $\alpha_{\rm eff}$ presents an eightfold ambiguity in the range $[0^{\circ},180^{\circ}]$. 
The eight solutions are 
$\alpha_{\rm eff}=(11 \pm 7)^{\circ}$, $\alpha_{\rm eff}=(41 \pm 7)^{\circ}$,
$\alpha_{\rm eff}=(49 \pm 7)^{\circ}$, $\alpha_{\rm eff}=(79 \pm 7)^{\circ}$, 
$\alpha_{\rm eff}=(101 \pm 7)^{\circ}$, $\alpha_{\rm eff}=(131 \pm 7)^{\circ}$,
$\alpha_{\rm eff}=(139 \pm 7)^{\circ}$, $\alpha_{\rm eff}=(169 \pm 7)^{\circ}$~\cite{ALPHA}.
Assuming that the relative strong phase between the relevant tree amplitudes is negligible \cite{BOUNDS} it is possible to reduce this ambiguity
to a twofold ambiguity in the range $[0^{\circ},180^{\circ}]$:
$\alpha_{\rm eff}=(11 \pm 7)^{\circ}$, $\alpha_{\rm eff}=(79 \pm 7)^{\circ}$.
BaBar combines the solution near $90^{\circ}$, 
$\alpha_{\rm eff}=(79 \pm 7)^{\circ}$, with the bounds on $|\alpha_{\rm eff}-\alpha|$ and estimates the 
weak phase $\alpha=(79 \pm 7 \pm 11)^{\circ}$. This solution is consistent with the current average value of $\alpha$, based on the 
analysis of $B\to \pi\pi$, $B\to \rho\rho$, and $B\to \rho\pi$ decays \cite{amsler,fitter}.

\section{Measurement of B decays to $a_1(1260)^{\pm}a_1(1260)^{\mp}$}
\noindent
Charmless $B$ decays to  final states involving two axial-vector mesons (AA)  have received considerable theoretical 
attention in the last few years \cite{Cheng, Calderon}. Using QCD factorization, the branching fractions of several  
$B\rightarrow AA$ decay modes have been calculated. Predictions for the branching fraction of the 
$B^0 \rightarrow a_1(1260)^{\pm}a_1(1260)^{\mp}$~\cite{notazione} decay mode vary between $37.4 \times 10^{-6} $ \cite{Cheng} and 
$6.4 \times 10^{-6} $ \cite{Calderon}. Branching fractions at this level should be observable with the BaBar data sample, 
which can be used to discriminate between the predictions. The predicted value of the longitudinal polarization fraction 
$f_L$ is 0.64 \cite{Cheng}. The only available experimental information on this $B$\ decay mode is the branching fraction 
upper limit (UL) of $2.8 \times 10^{-3}$\ at 90\% confidence level (CL) measured by CLEO \cite{a1a1CLEO}.
The measured value $f_L \sim 0.5$\ in penguin-dominated$B \rightarrow \phi K^*$ decays \cite{PhiKst} is in contrast with naive 
SM calculations predicting a  dominant longitudinal polarization ($f_L \sim 1$) in $B$ decays to vector-vector (VV) 
final  states. The naive SM expectation is confirmed in the tree-dominated $B \rightarrow \rho \rho$ \cite{rho,RhoRho,newrho} and 
$B^+ \rightarrow \omega \rho^+$ \cite{OmegaRho} decays.  A value of $f_L \sim 1$ is found in vector-tensor $B \rightarrow \phi
K^*_2(1430)$ decays \cite{Kst2}, while  $f_L \sim 0.5$ is found in $B \rightarrow \omega K^*_2(1430)$\ decays \cite{OmegaRho}
(see Ref~\cite{PDGReview} for further discussion). The small value of $f_L$ observed in $B \rightarrow \phi K^*$\ decays has stimulated
theoretical effort, such as the introduction of non-factorizable terms and penguin-annihilation amplitudes \cite{SMCal}. Other explanations 
invoke new physics \cite{NP}. Measurement of $f_L$ in $a_1 a_1$~\cite{notazione} decays will provide additional information. 
The analysis uses an integrated luminosity of $423.0~fb^{-1}$, corresponding to $(465 \pm 5) \times 10^6 B\bar{B}$ pairs, recorded at the $\Upsilon (4S)$ 
resonance~\cite{Paolo}. BaBar has measured the branching fraction:
${\cal B}(B^0 \rightarrow a_1^+ a_1^-) \times [{\cal B} ( a_1^+ \rightarrow (3\pi)^+ )]^2 = 11.8\pm2.6\pm1.6 \times 10^{-6}$
and the fraction of longitudinal polarization $f_L = 0.31 \pm 0.22 \pm  0.10$. Assuming that 
${\cal B}(a_1^+ \rightarrow \pi^- \pi^+ \pi^+)$ is equal to ${\cal B}(a_1^+ \rightarrow \pi^+ \pi^0 \pi^0)$, and that 
${\cal B}( a_1^+ \rightarrow (3\pi)^+)$ is equal to 100\%~\cite{amsler} the branching fraction is   
${\cal B}(B^0 \rightarrow a_1^+ a_1^-)$ = $(47.3 \pm 10.5 \pm 6.3)\times 10^{-6}$. The decay mode is seen 
with a significance of $5.0 \sigma$, which includes systematic uncertainties. The measured branching fraction and 
longitudinal polarization are in general agreement with the theoretical expectations in~\cite{Cheng}. In principle an indipendent measurement 
of $\alpha$ of the UT can be extracted from the time-dependent analysis of $CP$-violating asymmetries of $B^0 \rightarrow a_1(1260)^{\pm}a_1(1260)^{\mp}$
but the available statistics is indeed too low to perform such an analysis. Potentially a new generation of Super Flavor Factories ~\cite{belle,superb} 
will be able to achieve this result. 

\section{Summary}
\noindent
In summary significant progress in the measurement of $\alpha$ has been made over the last decade by BaBar. In $B\rightarrow \rho\rho$ decays recent
Babar results have substancially improved the precision of the measurement and at the same time a new measurement of $\alpha$ has been extracted from 
$B \rightarrow a_1(1260) \pi$ decays. All these results are in good agreement with the predictions obtained by SM-based fits~\cite{fitter}.


\begin{thebibliography}{99}
\bibitem{a1}
For the $a_1(1260)$ meson we use the short notation ${a_1}$.
\bibitem{notazione}
$a_1$\ will be used to indicate the $a_1(1260)$\ meson.
\end{thebibliography}
\end{document}